# The WHO surveillance threshold and the emergence of drug-resistant HIV strains in Botswana

Raffaele Vardavas[1,2] and Sally Blower[1,2]

[1] Department of Biomathematics and UCLA AIDS Institute, David Geffen School of Medicine at UCLA,

1100 Glendon Avenue PH2, Los Angeles, CA 90024, USA

[2] The Semel Institute for Neuroscience and Human Behavior.

**Correspondence** and request for material should be addressed to S.B. (sblower@mednet.ucla.edu).

**ABSTRACT**

**Background**

Approximately 40% of adults in Botswana are HIV-infected. The Botswana antiretroviral program began in 2002 and currently treats 34,000 patients with a goal of treating 85,000 patients (~30% of HIV-infected adults) by 2009. We predict the evolution of drug-resistant strains of HIV that may emerge as a consequence of this treatment program. We discuss the implications of our results for the World Health Organization's (WHO's) proposed surveillance system for detecting drug-resistant strains of HIV in Africa.

Methods

We use a mathematical model of the emergence of drug resistance. We incorporate demographic and treatment data to make specific predictions as to when the WHO surveillance threshold is likely to be exceeded.

Results

Our results show – even if rates of acquired resistance are high, but the drug-resistant strains that evolve are only half as transmissible as wild-type strains – that transmission of drug-resistant strains will remain low (< 5% by 2009) and are unlikely to exceed the WHO's surveillance threshold. However, our results show that transmission of drug-resistant strains in Botswana could increase to ~15% by 2009 if resistant strains are as transmissible as wild-type strains.

Conclusion

The WHO's surveillance system is designed to detect transmitted resistance that exceeds a threshold level of 5%. Whether this system will detect drug-resistant strains in Botswana by 2009 will depend upon the transmissibility of the strains that emerge. Our results imply that it could be many years before the WHO detects transmitted resistance in other sub-Saharan African countries with less ambitious treatment programs than Botswana.

**Background**

Antiretroviral therapy (ART) has drastically reduced mortality and morbidity rates in developed countries. However, patients may not fully adhere to their medications, and drug-resistant strains may evolve *in vivo*. Patients who develop acquired resistance may transmit these drug-resistant strains. In Europe and North America there are now fairly high levels of acquired, and transmitted, resistance [1]. Hence there is concern that usage of ART in Africa (which is just beginning) could quickly lead to the evolution of drug-resistant strains, and cause a high level of transmitted resistance. This may potentially reduce the effectiveness of control efforts. Thus, the World Health Organization (WHO) has developed a surveillance system [2, 3] for monitoring the emergence of transmitted resistance in Africa. Their system will not estimate the precise incidence of transmitted resistance. It is based upon a binomial sequential lot quality assurance sampling method and is designed to detect levels of transmitted resistance that reach (or exceed) a threshold value of 5% [2, 3]. Currently, it is unknown when this threshold is likely to be reached in any country, because no country-specific predictions have yet been made. Here, we predict the emergence of transmitted drug-resistant strains in Botswana; we include demographic and treatment data in a mathematical model. Specifically, we predicted the likely temporal dynamics of these strains by varying: (i) the rate of acquired resistance, and (ii) the relative transmissibility of drug-resistant strains. Then, we discuss the implications of our results in the context of the WHO's surveillance system.

Today, Botswana is one of the world's worst hit countries by the AIDS pandemic. An estimated 39% [4] of Botswana's 730,000 adults between the ages of 15 and 49 are HIV-infected. In 2002 HIV/AIDS was declared the most serious threat to the country, and in that same year their ART program was initiated. Currently the program has 34,000 patients on treatment [5]. The goal is to treat 85,000 patients (or equivalently 30% of HIV-infected individuals) by 2009 [6]. Botswana has by far the highest treatment rate in Africa, and has already slightly exceeded their scheduled treatment goals. Botswana is now considered a test nation by foreign investors for ART programs in sub-Saharan Africa. Recent substantial financial investments have enabled the opening of 31 treatment centers, ensuring that Botswana is likely to reach its treatment goals by 2009.

The first-line drug regimen used in Botswana is a combination of Zidovudine, Lamivudine, and Nevirapine or Efavirenz [7]. Under perfect adherence, only 5-10% of patients are likely to develop drug resistance in the first year [8]. However, if adherence is less than perfect, a greater proportion of patients will develop drug resistance within a year [9]. In Botswana the principal barriers to adherence include financial constraints, stigma, poor treatment accessibility, interruption of drug supply and toxicities [10]. Patients who acquire resistance have the potential to transmit drug-resistant strains via sexual transmission. Currently, relatively little is known about the transmissibility of drug-resistant strains of HIV *in vivo*. However, *in vitro* experiments have shown that the replication rate of drug-resistant strains is generally less than that of wild-type strains [11-13]. Thus, it is generally assumed that drug-resistant strains of HIV are less transmissible than wild-type strains.

**Methods**

To predict the temporal dynamics of transmitted resistance to 2009 we used a mathematical model. We modeled the dynamics of the emergence (by acquired resistance) and the transmission of drug-resistant strains as the result of ART. In the model, sexually active adults are classified into one of four states: (i) susceptible; (ii) infected with wild-type strains and treatment-naïve; (iii) infected with wild-type strains and on treatment; (iv) and infected with drug-resistant strains. Our model captures the essential processes of HIV transmission dynamics. Each year a proportion of the drug-naïve HIV-infected population can initiate treatment, and a proportion of the treated population can discontinue treatment [14]. Treatment benefits were modeled by increasing life expectancy and by reducing viral load (and hence transmissibility) in treated patients in comparison with untreated HIV-infected individuals. The population of treated individuals, has an average annual probability of developing drug resistance specified by the parameter *r*. Treated individuals that develop drug resistance increase their viral load and progress to AIDS faster than treated individuals infected with wild-type strains. Individuals infected with drug-resistant strains can transmit these strains; the transmissibility of drug-resistant strains is specified by the parameter $\beta_R$. Our model captures the dynamics of transmitted drug resistance that have occurred due to sexual transmission, we note that this is the type of transmitted resistance that will be monitored in Botswana by the WHO surveillance system. We did not include the process of vertical (mother to child) transmission of drug-resistant strains. We used demographic and treatment data from the Botswana ART program to predict the dynamics of transmitted drug-resistant strains to 2009.

The evolution of transmitted resistance is driven by two key parameters: the rate of development of acquired resistance ($r$) and the transmissibility of drug-resistant strains ($\beta_R$) [5, 15, 16]. We varied the values of these two parameters and made six predictions. The probability of acquired resistance depends upon adherence, viral replication rate, and the specific mutations that arise. Acquisition of one resistance mutation will not necessarily reduce the efficacy of a regimen. We modeled the average rate of development of acquired resistance in a group of patients rather than the rate of accumulation of resistant mutations in an individual patient. Thus, to make three years predictions, we assumed that the average percentage of treated individuals that will develop acquired resistance is approximately constant over the short-term. We assumed that on average either 20% or 33% of treated patients could develop resistance per year[17-21]. The relative transmissibility of resistant strains will depend upon which mutations are present. Some mutations will substantially reduce transmissibility whereas other mutations may have little effect. Hence, we examined a wide variety of mutations by varying the relative transmissibility from ~0% reduction to a 75% reduction in relative transmissibility of drug-resistant strains relative to wild-type strains. Specifically, we assumed that the drug-resistant strains that may evolve could be: (i) only 25% as transmissible, (ii) only 50% as transmissible, or (iii) equally as transmissible, as wild-type strains [12, 22].

We calculated the treatment rate from the projected number of patients that the Botswana ART program plans to treat over the next three years. The program plans to treat 85,000 patients by 2009 [6] (Figure 1). The treatment rate was found by assuming that it remains constant such that by the year 2009 Botswana will have 85,000 patients that have received ART. The comparison between our modeled treatment rate (dashed gray line) and the actual treatment data from 2002 to 2005 (dashed red line) is shown in Figure 1. Parameter values used in our model are given in Figure Legend 2.

**Results**

Our results show that whether the WHO surveillance threshold of 5% is exceeded (or not) by 2009 will be very dependent upon both the transmissibility/transmissibility of the drug-resistant strains that evolve and the rate of development of acquired resistance (Figure 2). Data shown in Figure 3 are average predictions, and the vertical bars show the expected probabilistic fluctuation range of the possible values within one standard deviation (Figure 2). Panels (A), (C) and (E) show probabilistic predictions assuming that 20% of treated patients acquire drug resistance per year; this rate implies that, on average, treated patients would develop resistance in 5 years. Panels (B), (D) and (F) show probabilistic predictions assuming 33% of treated patients acquire drug resistance per year; this implies that, on average, treated patients would develop resistance in 3 years. Transmissibility of drug-resistant strains relative to the wild-type strains is: (i) 25% in panels (A) and (B), (ii) 50% in panels (C) and (D), and (iii) 100% in panels (E) and (F).

Our results show that if the drug-resistant strains that evolve are only 25% as transmissible as the wild-type strains then transmitted resistance will reach at most just below 3% in the next three years (Figure 2B). Therefore, the WHO surveillance threshold is not likely to be exceeded by 2009, even if the rate of acquired resistance is very high (Figure 2B). However, our predictions show that transmitted drug-resistance would reach 6% by 2009 if the drug-resistant strains that evolve are 50% as transmissible as the wild-type strains and the rate of acquired resistance is very high (Figure 2D). Under these conditions, the WHO surveillance threshold is likely to be exceeded. If the drug-resistant strains that evolve are 50% as transmissible as the wild-type strains, but the rate of development of acquired resistance is only 20% per year, transmitted resistance would only rise to just below 5% by 2009 (Figure 2C). Thus, the WHO surveillance threshold may not be exceeded. However, our results show the expected probabilistic fluctuations are large enough that there is a fair probability that transmitted drug resistance could rise to just above 5% in the next three years, and thus exceed the surveillance threshold. Finally, if the drug-resistant strains that emerge are as transmissible as the wild-type strains then the WHO threshold will be reached in 2006 (Figure 2F) or 2007 (Figure 2E); under these conditions, levels of transmitted resistance could be as high as 13% by 2009 (Figure 2F). Obviously, if drug-resistant strains evolve that are more transmissible than wild-type strains the levels of transmitted resistance will be even higher (results not shown).

**Discussion**

The WHO surveillance scheme for detecting transmitted resistance uses a binomial sequential lot quality assurance sampling (LQAS) method [23, 24]. The LQAS methodology has been used by the WHO to determine a range for the minimum number of newly infected treatment-naïve adults that need to be sampled for surveillance. The specified minimum sample necessary to detect a 5% threshold of transmitted resistance ranges from 50 to 70 newly infected treatment-naïve adults [2, 3, 24]. The LQAS methodology is easy to implement and is cost-effective, but it does not yield a precise prevalence or incidence estimate. Instead the prevalence of transmitted drug resistance is classified as less than 5%, 5% to 15% and over 15%. Thus the surveillance scheme will only determine whether the specified minimum threshold of 5% has been exceeded, or not.

We have used a mathematical model to predict the dynamics of drug-resistant strains in Botswana for the next three years. Unlike a previously published model of HIV transmission and ART [16], our model provides further insight as it enables predictions for both the average level of transmitted drug resistance and the expected probabilistic fluctuations about these average values. Our model does not distinguish between untreated and treated individuals infected with drug-resistant strains as in [16], although the two models give very similar short-term predictions (over a seven year period from 2002 to 2009). This indicates that using weighted averages of infectiousness and disease progression rates, and grouping untreated and treated individuals infected with drug-resistant strains together is an appropriate assumption to make when generating probabilistic short-term predictions.

The model developed by Blower *et al.* [16] includes additional processes, such as cases of acquired resistance and/or transmitted resistant cases reverting to drug-sensitive cases over time. These processes are not included in our probabilistic model that we used to make short-term predictions; however, these processes are important to include for obtaining an accurate description of long-term dynamics. We note that in Botswana over the next four years: (i) the number of cases of transmitted resistance, and (ii) the number of individuals developing acquired resistance, giving up treatment, and converting to drug-sensitive cases, will be very small. In the next three years in Botswana the number of individuals who are likely to acquire super-infection is also likely to be small. Thus including the possibility of reversion and/or super-infection in our probabilistic model was not necessary, since we generated only short-term predictions. Hence, it was appropriate to use our probabilistic model to predict the level of transmitted resistance in Botswana over the next three years.

**Conclusion**

It is possible that HIV infected individuals will obtain ART outside the official treatment program. This could potentially increase levels of transmitted resistance. However, since Botswana plans to treat a very high percentage of HIV infected patients the number of individuals who will obtain drugs outside the program is likely to be small. Thus, any additional drug resistance generated by these individuals is also likely to be small. Botswana will have achieved high treatment rates by 2009. Our results imply that, unless the strains of drug-resistant HIV that evolve in Botswana are extremely transmissible, the WHO threshold for detection of transmitted resistance will not be exceeded by 2009. We suggest that WHO surveillance should be initiated in Botswana (and other African countries) only when transmitted resistance is expected to have exceeded the threshold. However, sentinel sites for surveillance should be used through out the roll-out [1]. The mathematical model that we have presented can be used to make predictions for transmitted resistance for other countries where the roll-out of ART is just beginning. Our results imply that it is likely that it will be many years before the WHO will detect transmitted drug resistance in other sub-Saharan African countries that have less ambitious treatment programs than Botswana.

**Acknowledgements.** We thank Erin Bodine, Romulus Breban, Tom Chou, James Kahn, Justin Okano and David Wilson for technical discussions. RV and SB are grateful for the financial support of NIH/NIAID (RO1 AI041935).

**Competing interest statement.** None.

Supplemental Material that includes mathematical derivations and equations used to make our predictions is available on request by contacting R.V. at vardavas@ucla.edu

**Figure Captions**

**Figure 1.** Empirical data from the Botswana treatment program are shown by the filled boxes. The aim of the Botswana ART program is to reach 85,000 patients by 2009 (black dashed line)[6]. The treatment used by our model uses the linear fit shown by the dashed gray line, this gives a constant per capita treatment rate of 0.050 per year.

**Figure 2.** Predictions from our model showing quarter yearly expected percentage values of newly infected treatment-naïve adults that are infected with drug-resistant strains. The bars represent the possible range of values due to the stochastic fluctuation contained in one standard deviation over the expected percentage. Here, we have assumed that the average progression time to AIDS for HIV infected patients that are treatment-naïve, treated drug-sensitive and drug-resistant are 10, 18 and 12 years respectively[22, 25]. The untreated and treated drug-sensitive transmissibility coefficient per partnership were set to 0.12 and 0.04 respectively[26-28]. Furthermore, we have assumed that every year one tenth of the treated drug-sensitive population suspend ART in Botswana[29]. The population of treated individuals, has an average annual probability of developing drug resistance specified by the parameter *r*, and the transmissibility of drug-resistant strains is specified by the parameter $\beta_R$. The six panels show predictions using the following parameter sets: **(A)** $r^{-1}=5$ years, $\beta_R = 0.03$ **(B)** $r^{-1}=3$ years, $\beta_R = 0.03$ **(C)** $r^{-1}=5$ years, $\beta_R = 0.06$ **(D)** $r^{-1}=3$ year, $\beta_R = 0.06$ **(E)** $r^{-1}=5$ years, $\beta_R = 0.12$ **(F)** $r^{-1}=3$ years, $\beta_R = 0.12$.

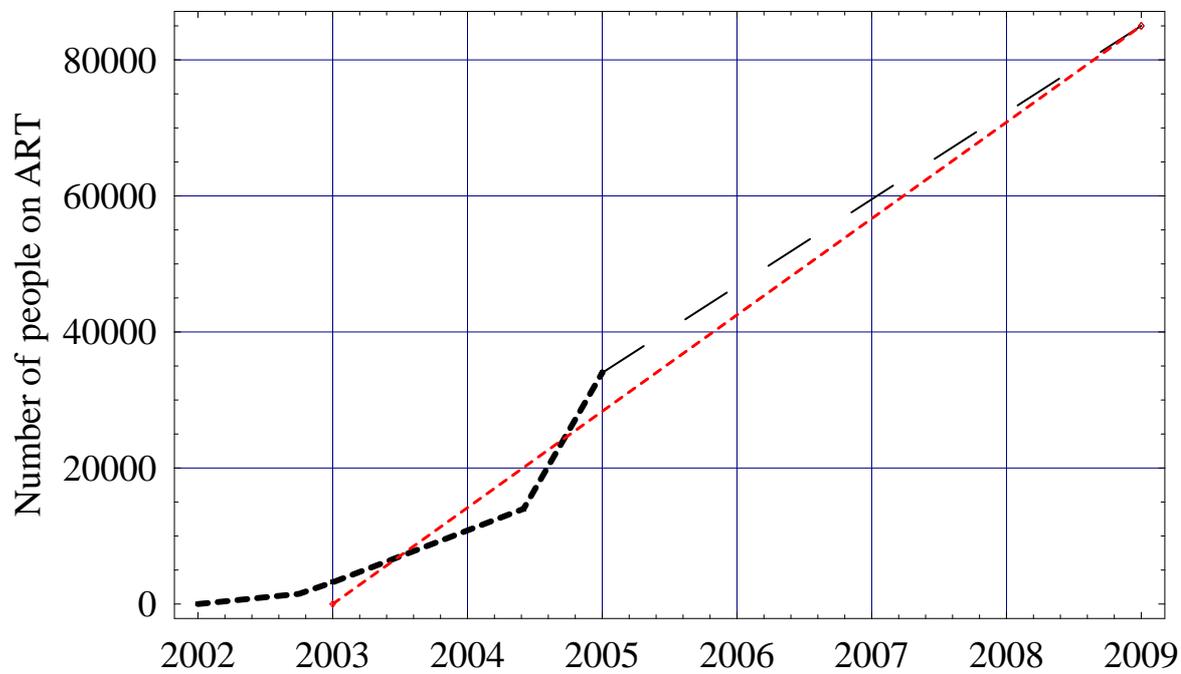

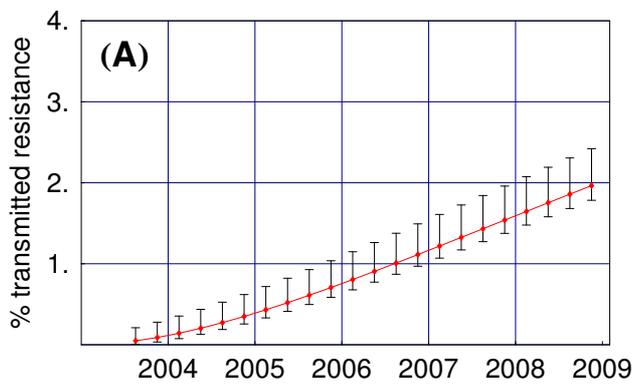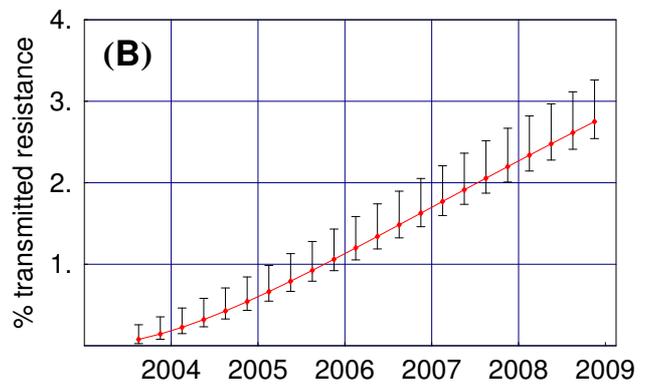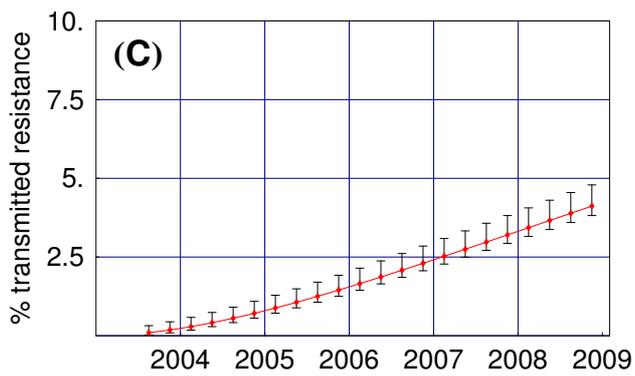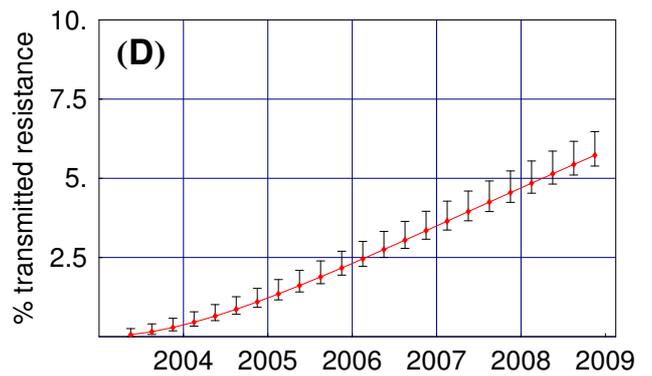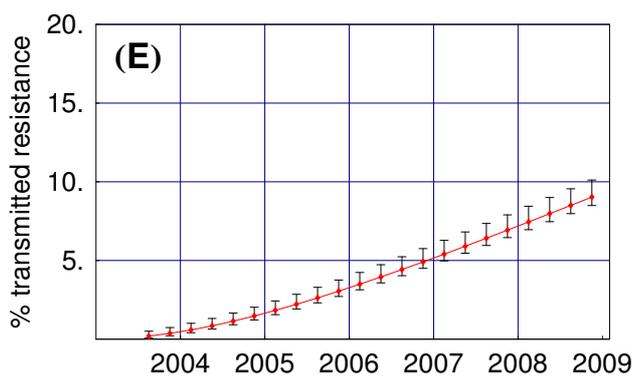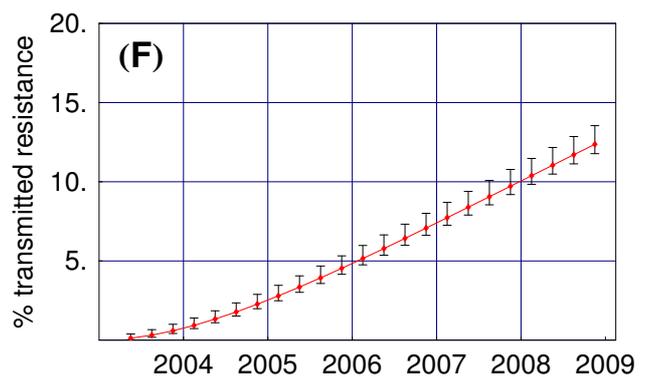